# Lexicographic Multi-objective Geometric Programming Problems

Dr.A. K. Ojha[1] and K. K. Biswal[2]

[1]School of Basic Sciences, IIT Bhubaneswar,
Orissa, Pin-751013, India

[2]Department of Mathematics, CTTC Bhubaneswar,
B-36, Chandaka Industrial Area, Bhubaneswar, Orissa, Pin-751024, India

**Abstract**
A Geometric programming (GP) is a type of mathematical problem characterized by objective and constraint functions that have a special form. Many methods have been developed to solve large scale engineering design GP problems. In this paper GP technique has been used to solve multi-objective GP problem as a vector optimization problem. The duality theory for lexicographic geometric programming has been developed to solve the problems with posynomial in objectives and constraints.

**Keywords**: *Lexicographic minimization, Geometric programming, duality theory, vector minimization and vector maximization.*

## 1. Introduction

In many real-life optimization problems, multiple objectives have been taken into account, which may be related to the economical, social and environmental aspects of optimization problems. The multiple objectives are usually incommensurate and in conflict with one another. In general, a multiple objective optimization problem does not have a single solution that could optimize all objectives simultaneously. It never search for optimal solution but for efficient solution that can best suited compromise solution to all multiple objectives. Though the geometric programming technique due to Duffin et al.[3] helps to solve various types of nonlinear single objective posynomial problems but there are very few work have been made in this direction to solve multiple objective GP problems. Lexicographic optimization approach is one such technique to handle multiple objective GP problem. Several mathematical and game theoretic applications of nonlinear lexicographic optimizations are reported by Behringer[1]. Impressed upon the work of Behringer, Nijkamp[6] applied the lexicographic optimization technique in a land use problem for industrial activities in a newly created industrial area in a Rhine-delta region near Rotterdam. Application of linear lexicographic due to Isermann[4] and Turnovec[8] strengthen to handle scalar valued optimization problems. Biswal[2] used fuzzy programming[10] to solve multi-objective geometric problem where as lexicographic order and duality has been studied by Martinez[5]. In this paper we have applied lexicographic geometric programming technique to solve special type of multi-objective optimization problem.

The organization of the paper is as follows: Following introduction the definition of multi-objective geometric programming and lexicographic optimization have been discussed in Section-2 and 3 respectively. Definition of lexicographic geometric programming has been discussed in Section 4 and the numerical examples have been incorporated in Section 5. Finally the conclusion has been presented in Section 6.

## 2. Multi-objective geometric programming

A multi-objective geometric programming problem can be defined as:
Find x = (x$_1$, x$_2$,… ,x$_n$)$^T$ so as to

$$\min : g_{k0}(x) = \sum_{t=1}^{T_{ko}} C_{k0t} \prod_{j=1}^{n} x_j^{a_{k0tj}}, k = 1,2,...,p$$

(2.1)

subject to

$$g_i(x) = \sum_{t=1}^{T_i} C_{it} \prod_{j=1}^{n} x_j^{d_{itj}} \leq 1, i = 1,2,...,m \quad (2.2)$$

$$x_j > 0, j = 1,2,...,n \quad (2.3)$$

where $C_{k0t}$ for all k and t are positive real numbers and $d_{itj}$ and $a_{k0tj}$ are real numbers for all i, k, t, j.
$T_{k0}$ = number of terms present in the $k^{th}$ objective function.
$T_i$ = number of terms present in the $i^{th}$ constraint.
In the above multi-objective geometric programming problem there are p number of minimization type objective functions, m number of inequality type constraints and n number of strictly positive decision variables.

Let us define $F(x) = \{g_{10}(x), g_{20}(x),...., g_{p0}(x)\}$
and $G(x) = \{g_1(x) - 1, g_2(x) - 1,...., g_m(x) - 1\}$

Now the above optimization problem can be rewritten as:

$$lex \min: \{F(x): x \in R^n, G(x) \leq 0^m\} \quad (2.4)$$

which is called lexicographic geometric programming (LGP) problem.

## 3. Lexicographic optimization problem

Let us denote $O^n$ an n-dimensional zero vector and $O^{m \times n}$ an m×n zero matrix. An inequality of the type $x \geq O^n$ means $x \geq O^n$, but $x \neq O^n$

A vector $x \in R^n$ is said to be lexicographically non negative if either $x = O^n$ or its first non-zero component is positive and we denote it by $lex\, x \geq O^n$

Similarly a non zero vector $x \in R^n$ is said to be lexicographically non positive if its first non zero component is negative and it is denoted by $lex\, x < O^n$

An m×n matrix A is called lexicographically non negative if all its columns are lexicographically non negative and we denote it as

$$lex\, A \geq O^{m \times n}$$

In the similar manner we can define lexicographically non positive and lexico-graphically negative vector and its corresponding matrix.

Let F be an p-dimensional vector valued function defined on $R^n$ and $X \subset R^n$.

A vector $x^* \in X$ is said to be a lexicographically minimal point of F with respect to X if for any $x \in X$ such that $lex\, F(x^*) \leq F(x)$

The problem of finding lexicographically minimum point of F with respect to X is called lexicographically minimizing problem which is denoted as:

$$lex \min: \{F(x): x \in X\} \quad (3.1)$$

Similarly a point $\bar{x} \in X$ is said to be lexicographically maximum point of F with respect to X if $lex\, F(\bar{x}) \geq F(x)$ and this problem is called lexicographically maximizing problem given by

$$lex \max: \{F(x): x \in X\} \quad (3.2)$$

If we assume

$$X = \{x \in R^n: G(x) \leq O^m\} \quad (3.3)$$

then the lexicographically minimizing problem can be defined as,

$$lex \min: \{F(x): G(x) \leq O^m\} \quad (3.4)$$

A lexicographically minimum point of F with respect to (3.3) is called an optimal solution to the problem given by (3.4).

The problem given by (3.4) is called convex optimization if all the components of F and G are convex functions.

## 4. Lexicographic geometric Programming

The lexicographic multi-objective geometric programming defined by

$$lex \min: \{F(x): x \in R^n, G(x) \leq O^m\}$$

where the functions are defined by (2.1), (2.2) and (2.3). Now we will prove the following theorem for the existence of unique optimal solution of the lexicographic optimization GP problem.

**Theorem:-** If the primal problem of the geometric programming is consistent and its dual program has a maximizing point with strictly positive components then the primal problem of geometric programming has a unique optimal solution if and only if the rank of its exponent matrix is equal to the number of columns.

**Proof :-** From the duality theorem due to Duffin[3], we know that for each optimization point $x^*$ for the primal problem there exist a maximizing point $w^*$ of dual program which is given by the following equations.

$$C_{k0t} \prod_{j=1}^{n} x_j^{a_{k0tj}} = w_{k0t}^* v(w^*), k = 1,2,...p, t = 1,2,....T_{k0}. \quad (4.1)$$

$$C_{it} \prod_{j=1}^{n} x_j^{d_{itj}} = \frac{w_{it}^*}{\lambda_i(w^*)}, i=1,2,...m, t=1,2,...T_i. \quad (4.2)$$

Taking logarithm on both sides of equation (4.1) and (4.2) we have

$$\ln(x_1^{a_{k0t1}} x_2^{a_{k0t2}} ... x_n^{a_{k0tn}}) = \ln\frac{w_{k0t} v(w^*)}{C_{k0t}}, k=1,2,...p, t=1,2,...T_{k0}$$

$$\ln(x_1^{d_{it1}} x_2^{a_{it2}} ... x_n^{a_{itn}}) = \ln\frac{w_{it}^*}{C_{it}\lambda_i(w^*)}, i=1,2,...m, t=1,2,...T_i.$$

which can be expressed as

$$a_{k0t1}\ln x_1 + a_{k0t2}\ln x_2 + .... + a_{k0tn}\ln x_n = \ln(\beta_{k0t});$$
$$k=1,2,....p; \; t=1,2,...,T_{k0} \quad (4.3)$$

And

$$d_{it1}\ln x_1 + d_{it2}\ln x_2 + .... + d_{itn}\ln x_n = \ln(\beta_{it});$$
$$i=1,2,....m; t=1,2,...T_i \quad (4.4)$$

where $\beta_{k0t} = \frac{w_{kot}^* v(w^*)}{C_{k0t}}, t=1,2,...T_{k0}$

and

$$\beta_{it} = \frac{w_{it}^*}{C_{it}\lambda_i(w^*)}, \; t=1,2,...T_i, i=1,2,....,m.$$

Now by substituting
$z_j = \ln x_j$ ; $j = 1, 2,...., n$

$$\ln(\beta_{k0t}) = \gamma_{k0t}, k=1,2,....p, \; t=1,2,......,T_{k0}$$

and $\ln \beta_{it} = \gamma_{it}, i=1,2,....m, t=1,2,....,T_i$

the equations (4.3) and (4.4) reduces to

$$a_{k0t1}z_1 + a_{k0t2}z_2 + ... + a_{k0tn}z_n = \gamma_{k0t}, \; t=1,2,...,T_{k0} \quad (4.5)$$

and

$$d_{it1}z_1 + d_{it2}z_2 + ... + d_{itn}z_n = \gamma_{it}, \; i=1,2,...m, t=1,2,...T_i \quad (4.6)$$

If we assume $T = \sum_{i=1}^{m} T_i$ then we will have T number of equations and n variables, which is exactly the dimension of the exponent matrix.
Writing the equations in the matrix form we have

$$\sum_{t=1}^{T_{k0}} w_{k0t} a_{k0tj} + \sum_{i=1}^{m}\sum_{t=1}^{T_i} w_{kit} d_{itj} = 0, \; k=1,2,...,p, j=1,2,...,n, i=1,2,...,m$$

$$Az = \gamma \quad (4.7)$$

where A is the exponent matrix of dimension T×n and T ≥ n. From the basic knowledge of linear algebra the system of equations (4.7) has a unique solution if and only if the rank of the matrix A is equal to the number of its columns. With the substitution

$$e^{z_j} = x_j \text{ or } z_j = \ln x_j, \; j=1,2,...,n$$

Our lexicographic optimization problem can be expressed as.

$$lex \min: \{F(z) : G(z) \leq 1_m\} \quad (4.8)$$

where F(z) is an p-dimensional vector valued function with

$$g_{k0}(z) = \sum_{t=1}^{T_{k0}} C_{k0t} e^{\sum_{j=1}^{n} a_{k0tj} z_j}, \text{ k=1,2,....,p}$$

and $g_i(z) = \sum_{t=1}^{T_i} C_{it} e^{\sum_{j=1}^{n} d_{itj} z_j}$, i=1,2,....,m

Due to the monotonicity of logarithm function the problem (4.8) can be expressed as:

$$lex \min: \{\ln F(z) : \ln G(z) \leq O^m\} \quad (4.9)$$

with

$$\ln F(z) = \{\ln g_{10}(z), \ln g_{20}(z),...., \ln g_{p0}(z)\}$$

and $\ln G(z) = \{\ln g_1(z), \ln g_2(z),...., \ln g_m(z)\}$

Introducing new variables

$$w_{kot} = \frac{C_{k0t} e^{\sum_{j=1}^{n} a_{k0tj} z_j}}{\sum_{t=1}^{T_{k0}} C_{k0t} e^{\sum_{j=1}^{n} a_{k0tj} z_j}}, \text{ k=1,2,...,p} \quad (4.10)$$

And $w_{kit} = u_{ki} \frac{C_{it} e^{\sum_{j=1}^{n} d_{itj} z_j}}{\sum_{t=1}^{T_i} C_{it} e^{\sum_{j=1}^{n} d_{itj} z_j}}$, k=1,2,...,p, i= 1,2,...,m  (4.11)

After suitable transformation [3] the dual problem associated with the lexicographic geometric programming problem can be obtained.

$lex \max : V(w)$

such that $\sum_{t=1}^{T_{k0}} w_{k0t} = 1$, k=1,2,....,p  (4.12)

IJCSI International Journal of Computer Science Issues, Vol. 6, No. 2, 2009
ISSN (Online): 1694-0784
ISSN (Print): 1694-0814

23

(4.13)

$$u_{ki} = \sum_{i=1}^{T_i} w_{kit}, \quad k=1,2,\ldots,p, \quad i=1,2,\ldots,m \quad (4.14)$$

where $V_k(w)$ is the p-dimensional vector valued function of the form

$$V_k(w) = \sum_{t=1}^{T_{k0}} \left(\frac{C_{k0t}}{w_{k0t}}\right)^{k0t} \prod_{i=1}^{m} \prod_{t=1}^{T_i} \left(\frac{d_{it}}{w_{kit}}\right)^{w_{kit}} \prod_{i=1}^{m} u_{ki}^{u_{ki}},$$

k=1,2,….,p                (4.15)

According to the theory of geometric programming problem (4.12) is called the normality conditions, (4.13) is called orthogonality condition and (4.15) is the dual function.

The number $d = \sum_{k=1}^{p} T_{k0} + \sum_{k=1}^{m} T_i - n - 1$ is the degree of difficulty.

As the problem with zero degree of difficulty is easily solvable then the dual problem can be solved to get the maximizing vector $w^*$. Since the vector $w^*$ is the unique solution to the dual constraints, it is also the maximizing vector for the dual problem.

Using this dual optimizing vector the optimal solution $x^*$ to the primal problem can be determined by using following relationships.

$$C_{k0t} x_1^{a_{k0t1}} x_2^{a_{k0t2}} \ldots x_n^{a_{k0tm}} = w_{k0t}^* V_k(w_{k0t}^*),$$

k=1,2,….,p                (4.16)

$$C_{it} x_1^{d_{it1}} x_2^{d_{it2}} \ldots x_n^{d_{itm}} = \frac{w_{it}^*}{u_{ki}^*}, \quad k=1,2,\ldots,p; i=1,2,\ldots,m$$

(4.17)

where $u_{ki}^* = \sum_{i=1}^{T_i} w_{kit}^*$, k=1,2,…,p     (4.18)

## 5. Numerical Example

**Example:** Let us consider a numerical example

$$lex \min F(x) = \{g_{10}(x) = x_1^{-1} x_2^{-1} x_3^{-2}, g_{20}(x) = x_1^{-1} x_2^{-3} x_3^{-5} + x_1^{-1} x_2^{-1}\}$$

subject to  $x_1 x_2 x_3^2 + x_2 x_3 \leq 10$

$x_1 x_3 \leq 2$

$x_1, x_2, x_3 > 0$

**Solution Procedure of LGP problems.**
**Step 1:**
At first the objective functions of the multi-objective GP problem are ranked according their priority. Let us assume that the first objective function is in priority one i.e. $P_1$ and the second objective function in priority 2 i.e. $P_2$ and so on, and $p^{th}$ objective function in priority p i.e $P_p$.

**Step 2:**
Then first objective function $g_{10}(x)$ is minimized subject to all the original constraints. Let the minimum of the first objective be $g_{10}^{(1)}$ at $x^{(1)}$. Then we move to step 3.

**Step 3:**
Then the second objective function $g_{20}(x)$ is minimized subject to the original constraints with one additional constraint, i.e. $g_{10}(x) \geq g_{10}^{(1)}$

Let the minimum value of the second objective function be $g_{20}^{(2)}$ at $x^{(2)}$. Then we move to next step.

**Step 4:**
In step 4 third objective function $g_{30}(x)$ is minimized subject to the original constraint with two additional constraints, i.e. $g_{10}(x) \geq g_{10}^{(1)}$

$$g_{20}(x) \geq g_{20}^{(2)}$$

Same procedure is repeated for all the objective functions.

**Step 5:**
Finally, the last objective function is minimized subject to all the original constraints plus additional p-1 constraints.

$g_{10}(x) \geq g_{10}^{(1)}, \quad g_{20}(x) \geq g_{20}^{(2)}, \quad \ldots g_{p-1,0}(x) \geq g_{p-1,0}^{(p-1)}$

Let A be the exponent matrix.

$$A = \begin{pmatrix} -1 & -1 & -2 \\ 1 & 1 & 2 \\ 0 & 1 & 1 \\ 1 & 0 & 1 \end{pmatrix}$$  The rank of the matrix is 2

which is less than the number of columns.
The dual program of the function $g_{10}(x)$ is as

$$\max: V(w) = \left(\frac{1}{w_{101}}\right)^{w_{101}} \left(\frac{1/10}{w_{111}}\right)^{w_{111}} \left(\frac{1/10}{w_{112}}\right)^{w_{112}} \left(\frac{0.5}{w_{211}}\right)^{w_{211}} (w_{111}+w_{112})^{w_{111}+w_{112}}$$

Subject to: $w_{101} = 1$

$-w_{101} + w_{111} + w_{211} = 0$

$-w_{101} + w_{111} + w_{112} = 0$

$-w_{101} + 2w_{111} + w_{112} + w_{211} = 0$

The solution of the dual program gives $w_{101} = 1$, $w_{111} = 0.666...$, $w_{112} = 0.3333334$, $w_{211} = 0.3333334$ and the mean value of dual $v(w^*) = 0.15$. Using the primal dual relationship we have the optimal solution of $g_{10}(x)$ are $x_1^* = 0.9086967$, $x_2^* = 1.514494$, $x_3^* = 2.200954$ and its primal optimal solution is 0.15.

Similarly the dual program of $g_{20}(x)$ is defined as,

$$\max: V(w) = \left(\frac{1}{w_{201}}\right)^{w_{201}} \left(\frac{1}{w_{202}}\right)^{w_{202}} \left(\frac{1/10}{w_{211}}\right)^{w_{211}} \left(\frac{1/10}{w_{212}}\right)^{w_{212}} \left(\frac{0.5}{w_{211}}\right)^{w_{211}} (w_{211}+w_{212})^{w_{211}+w_{212}}$$

Subject to: $w_{201} + w_{202} = 1$

$-w_{201} - w_{202} + w_{211} + w_{221} = 0$

$-3w_{201} - w_{202} + w_{211} + w_{221} = 0$

$-5w_{201} + 2w_{211} + w_{212} + w_{221} = 0$

The solution of the problem gives its optimal value $V_2(w^*) = 0.4316470 \times 10^{-1}$

with $w_{201}^* = 0.6666667$, $w_{202}^* = 0.3333...$, $w_{211}^* = 1$, $w_{212}^* = 1.3333...$, $w_{221}^* = 0$

Using primal dual relationship we have $x_1 = 3.020273$, $x_2 = 23.01163$, $x_3 = 0.2483217$

## 6. Conclusions

In this chapter solution procedure of lexicographic GP has been presented. Unless the objective functions ranked properly, lexicographic solution may not be acceptable to a design engineer. If there are p number of objective functions one may formulate p! no of ways priority, which is a very difficult task for a design engineer. Also sometimes more than one objective functions remain in a priority. Unless the priority is proper solution of a real life problem gives some abnormal result. To set the proper ranking, method of Analytic Hierarchy Process (AHP) may be adopted. Two popular method of AHP by Saaty[7] namely row-column Addpotion method and Eigen-value method is used to find the proper ranking (weights) of the objective function. If the weights of the objective function can be estimated, then using the weights multi-objective GP problem can be converted to a single objective GP problem and solved.